\documentstyle[12pt,epsf,colordvi]{article}
\textBlack
\Large
\begin{document}
\baselineskip 18pt
\begin{flushright}
\Black{{\bf Fermilab-CONF-95/290-E}}\\
\Black{August 4, 1995}
\end{flushright}
\setlength{\unitlength}{1cm}
\thicklines
\vspace{1.0cm}
\large
\begin{center}
{\bf SEARCHING FOR ANTIPROTON DECAY AT THE FERMILAB ANTIPROTON
ACCUMULATOR}
\footnote{Presented at the International Europhysics Conference on
High Energy Physics, Brussels, Belgium, 27 July - 2 August 1995} \\
\vspace{1.0cm}
{\bf S. Geer}\\
\vspace{0.1cm}
\normalsize
Fermi National Accelerator Laboratory, Batavia,
Illinois 60510, USA\\
\vspace{1.0cm}
Presented on behalf of the APEX Collaboration\\
\end{center}
\vspace{1.0cm}
\begin{center}
{\bf Abstract}
\end{center}
\normalsize
This paper describes an experimental search for antiproton decay at the
Fermilab Antiproton Accumulator. The E868 (APEX) experimental setup is
described. The APEX data
is expected to be sensitive to antiproton decay if the antiproton lifetimes is
less than a few times 100 000 years.
\vspace{1.0cm}

\section{Introduction}
There has been a considerable experimental effort devoted to the search for
proton decay. As a result we know that the proton lifetime
$\tau_p \geq O(10^{32})$~years. The CPT theorem requires that the proton and
$\overline{p}$ lifetimes are equal. A search for $\overline{p}$ decay with a
short lifetime ($\tau_{\overline{p}} < \tau_p$) tests the CPT theorem and the
intrinsic stability of antimatter. Although we have no deep theoretical
reason to suspect CPT violation we note that $\overline{p}$ decay with a short
lifetime would provide a natural although unconventional explanation for
baryogenesis.

In the past there have been a number of experimental searches for the decay of
antiprotons stored in ion traps or storage rings. The most sensitive results
to date are listed in Table~\ref{limits_table}.
In particular, the APEX test experiment
(T861) at the Fermilab Antiproton Accumulator achieved a sensitivity of order
1000 years for some decay modes \cite{t861_prl}.
Experience gained in the T861 experiment was
used to design an upgraded experiment (APEX) with the goal of improving on
the T861 sensitivity by about a factor of 1000 and extending the search to
include many additional possible decay modes. The APEX experiment was
proposed in September 1992, approved in spring 1993, installed in March
1995, and took data between April and July 1995.

\begin{table}\begin{center}\caption{Published experimental limits on lifetime /
branching ratio for antiproton decay.}
\vspace{0.5cm}
\begin{tabular}{c|l|r|l} \hline\hline
 & & & \\
Experiment & Mode & $\tau$/BR Limit & Reference \\ \hline
& & & \\
Ion Trap & inclusive & 3.4 months & Galbrielse et al.
\cite{galbrielse} \\
& & & \\
T861     & $\overline{p} \rightarrow e^-\gamma$    & 1848 years & Geer et al.
\cite{t861_prl} \\
         & $\overline{p} \rightarrow e^-\pi^0$     & 554 years & Geer et al.
\cite{t861_prl} \\
         & $\overline{p} \rightarrow e^-\eta$      & 171 years & Geer et al.
\cite{t861_prl} \\
         & $\overline{p} \rightarrow e^-K^{0}_{S}$ & 29 years & Geer et al.
\cite{t861_prl} \\
         & $\overline{p} \rightarrow e^-K^{0}_{L}$ & 9 years & Geer et al.
\cite{t861_prl} \\
& & & \\
\hline \hline
\end{tabular}
\label{limits_table}
\end{center}
\end{table}

\section{The APEX Experimental Setup}

The APEX experiment took data at the Fermilab Antiproton Accumulator
operating at 8.9 GeV/c with typical beam currents of
100~mA (corresponding to $10^{12}$ stored
antiprotons). The detector, which was located downstream of a 15.9~m
straight section in the 474~m circumference accumulator ring, was designed to
search for two-body decay modes with an electron in the final state.
The experimental setup (see Fig.~\ref{detector_fig}) consists of:
\setlength{\unitlength}{0.7mm}
\begin{figure}[hbt]
\vspace{-2.5cm}
\begin{picture}(150,155)(20,1)
\hspace{2.0cm}\mbox{\epsfxsize15.0cm\epsffile{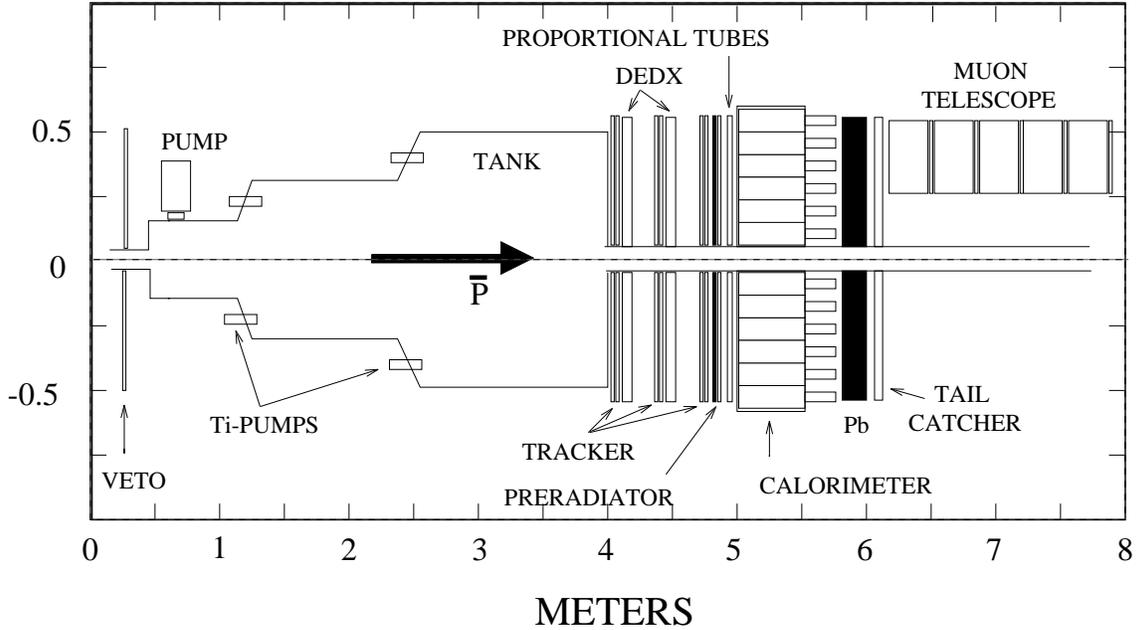}}
\end{picture}
\vspace{0.5cm}
\caption{Schematic of the APEX detector.}
\label{detector_fig}
\end{figure}
\begin{itemize}
\item[(i)] Decay tank. A 3.7~m long evacuated decay tank operated at
$10^{-11}$~Torr. The downstream end of the tank is a 1~m diameter cylinder
with a thin stainless steel window.
\item[(ii)] Veto counters. Upstream of the tank, 4 horizontal and 4 vertical
scintillation counters are arranged around the beampipe in a
$1 \times 1$~m$^2$ plane to veto tracks from upstream interactions.
\item[(iii)] Electromagnetic calorimeter. Lead-scintillator sampling
calorimeter consisting of 144 rectangular modules arranged in a $13 \times 13$
array with the 6 modules at each of the 4 corners missing, and the central
module absent to enable passage of the beam pipe. The modules are 17.7
radiation lengths deep and have transverse dimensions of
$10 \times 10$~cm$^2$.
Further details can be found in ref.~\cite{fcal}.
\item[(iv)] Scintillating fiber tracker. Between the tank window and the
calorimeter there are three vertical and three horizontal planes of 2~mm
scintillating fibers. Each plane consists of 768 fibers read out by 4
Hamamatsu multianode photomultipliers.
\item[(v)] dE/dx and preradiator counters. Between the tank window and the
calorimeter there are three planes of scintillation counters. The first two
planes are designed to measure the dE/dx of traversing charged particles, and
determine how many charged particles are in the event. The third plane is
downstream of a 2 radiation lengths thick lead radiator, and is designed
to assist in electron-hadron discrimination. Each scintillator plane is
constructed from four $1.3 \times 50 \times 100$~cm$^3$ counters read out
at both ends. Two of the counters are hung vertically on either side of
the beampipe, and two counters are hung horizontally, one above and one
below the beampipe.
\item[(vi)] Proportional tubes. Downstream of the preradiator there are
56 horizontal and 56 vertical proportion tubes designed to provide position
information for photon showers. The tubes have a $1 \times 1$~square inch
cross-section.
\item[(vii)] Tail catcher to aid electron-hadron discrimination.
Downstream of the calorimeter there is a 20~cm deep lead wall
followed by horizontal and
vertical planes of scintillation counters.
\item[(viii)] Muon telescope. Finally, downstream of the tail catcher a
$30 \times 30$~cm$^2$ telescope provides muon identification within a
limited acceptance. The telescope consists of 5 modules, each module having
30~cm of iron absorber followed by a scintillation counter.
\end{itemize}
\begin{figure}[hbt]
\vspace{4.0cm}
\begin{picture}(150,155)(0,1)
\mbox{\epsfxsize15.0cm\epsffile{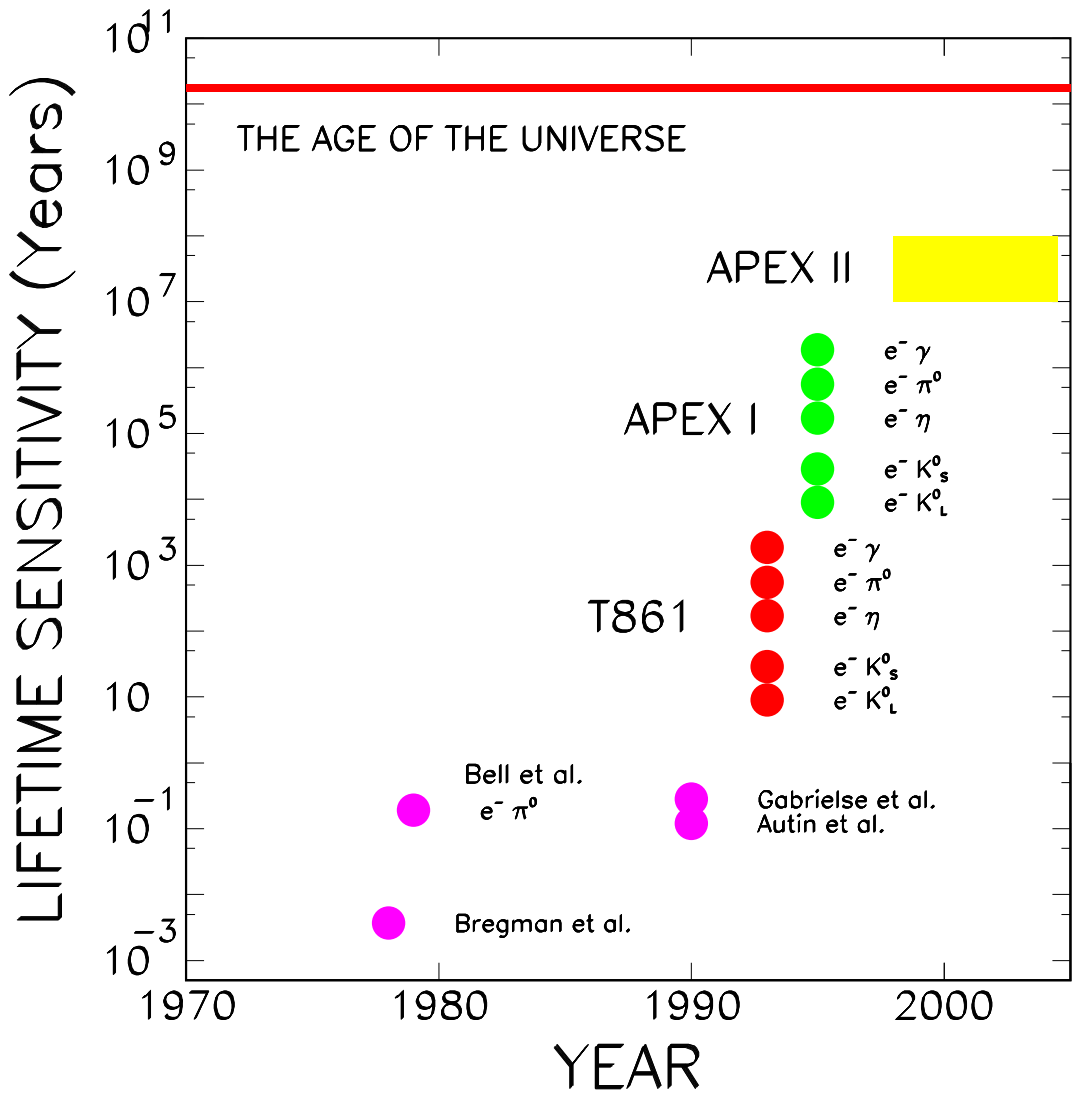}}
\end{picture}
\vspace{0.5cm}
\caption{Past experimental results, present expectations (APEX), and
future hopes (APEX II).}
\label{year_fig}
\end{figure}
\section{Data Sample and Future Prospects}
The APEX data sample is currently being analysed.
A measure of the potential sensitivity of the experiment is given by
the antiproton current in the accumulator integrated over the data taking time.
The APEX data sample corresponds to 26 000~mA-hours. This
means that if the $\overline{p}$ lifetime was $2 \times 10^9$~years then on
average one $\overline{p}$ should have decayed somewhere in the accumulator
during our data taking. To obtain the sensitivity of the data sample we need
to know the overall efficiency of the experiment. The decay tank occupies
about 1\% of the accumulator ring. Other geometrical factors together with
trigger and selection efficiencies are expected to result in an overall
efficiency of about 0.1\%. Hence the single event sensitivity of the data
sample is expected to be of order $10^6$~years.

The results of past $\overline{p}$ decay searches, present expectations,
and future hopes are summarized in Fig.~\ref{year_fig}.
In the future we can imagine a second generation experiment (APEX II) with a
single event sensitivity of order $10^8$~years, and a correspondingly improved
background rejection. At present there are plans to upgrade the Fermilab
antiproton accumulator that may or may not be compatible with a future APEX II.
If APEX II proves to be practical then in a few years we might be
able to find or exclude antiproton decay with lifetimes up to about 1\% of the
age of the universe !
\setcounter{secnumdepth}{0} 
%
\section{Acknowledgments}
The APEX experiment is supported by the U.S. Dept. of Energy and the National
Science Foundation.
%
%

%
\end{document}